\documentclass{elsart}

\usepackage{graphicx,amssymb,latexsym,color}

\journal{Physica B}

\begin{document}

\begin{frontmatter}

\title{Magnetoconductivity of Hubbard bands induced in Silicon MOSFETs}

\author{T. Ferrus\thanksref{email}},
\author{R. George},
\author{C.H.W. Barnes},
\author{N. Lumpkin},
\author{D.J. Paul},
\author{M. Pepper}

\thanks[email]{E-mail : taf25@cam.ac.uk}

\address{Cavendish Laboratory, University of Cambridge, J. J. Thomson Avenue, CB3 0HE, Cambridge, United Kingdom}

\begin{abstract}

Sodium impurities are diffused electrically to the oxide-semiconductor interface of a silicon MOSFET to create an impurity band. At low temperature and at low electron density, the band is split into an upper and a lower sections under the influence of Coulomb interactions. We used magnetoconductivity measurements to provide evidence for the existence of Hubbard bands and determine the nature of the states in each band.

\end{abstract}

\begin{keyword}
Silicon MOSFET, sodium, magnetoconductivity, localization, hopping, interference
\PACS 71.23.Cq \sep 71.55.Gs \sep 71.55.Jv \sep 72.15.Rn \sep 72.20.Ee \sep 72.80.Ng \sep 73.20.At \sep 73.40.Qv
\end{keyword}
                         
\end{frontmatter}

\section{Introduction}

In the early 70's a substantial effort was put into the study of the effect of impurities in silicon MOSFETs. The main reason for this was to understand the origin of the instabilities in field effect transistors and the umpredictable changes in the threshold voltages. These studies contributed to improve the performance of a new generation of silicon-based chips for computer applications and their reliability. In fact, the presence of impurities or defects in the silicon oxide leads to an inhomogeneity in the electron distribution at the Si-SiO$_2$ interface and traps electrons. The most common impurities are alkali metals like lithium, potasium or sodium. These trap states give rise to the formation of an impurity band below the conduction band. This effect was first observed by Fang, Hartstein and Pepper \cite{Fang, Hartstein, Pepper} for high impurity concentrations ($>1\times 10^{12}$\,cm$^{-2}$). A consequence of the existence of such an impurity band is that the onset voltage for conduction in oxide-doped MOSFETs is also shifted and the electron mobility substantially decreased. The impurity band produced by such a doping can be described by tight-binding models for high impurity density \cite{Suton} whereas Mott-Hubbard models \cite{Hubbard} need to be used for the low concentrations for which Coulomb interaction plays a defining role. Given single valency atoms like sodium, it would be expected that the electronic states in the band are made of single trapped electrons. Under the influence of electron-electron interactions, it has been suggested however that a stable state with two bound electrons would exist and would be characterized by a long lifetime (D$_-$ state) \cite{Norton, Taniguchi}. The band formed by the D$_-$ states is referred as the upper Hubbard band (UHB) and the one formed with neutral states as the lower Hubbard band (LHB). The question of the existence of Hubbard bands in Silicon MOSFETs has been put forward by Mott nearly thirty years ago to explain the magnetoconductivity of short disordered MOSFETs but the bands were never directly observed in experiments \cite {UHB}. The study of Hubbard bands in semiconductors has regained some attention since the end of the 90's with the development of quantum information and quantum computation \cite{Kane}. Following the difficulty in reading out directly the value of the spin in architectures developed in agreement with the Kane model, a spin to charge conversion was proposed. In this approach a stable D$_-$ state is used to read out the result of the quantum operations. Some optical studies have previously been carried out in Mott-Hubbard insulators like Sr$_2$CuO$_3$ \cite{Kishida} or boron-doped diamond \cite{Wu} but no direct conductivity measurement has been performed so far. In the present paper, we have used magnetoconductivity measurement to provide direct evidence of the presence of Hubbard bands in a sodium-doped silicon MOSFET. We also describe the electronic configuration of the obtained D$_-$ state by distinguishing between 1s$^2$, 1s2s and 1s2p states.

\section{Experiment}

\begin{figure}
\begin{center}
\includegraphics*[width=25pc]{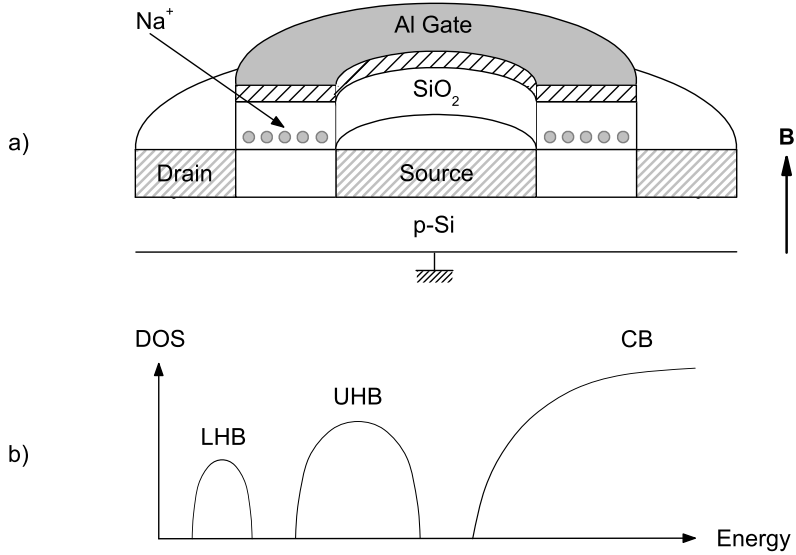}
\end{center}
\caption{a) Cross-section view of a corbino MOSFET used in the experiment when the sodium ions are close to the Si-SiO$_2$ interface. b) Schematic diagram of the density of states (DOS) for the present device, with a low energy (LHB) and a high energy (UHB) impurity band separated by a gap to the conduction band (CB).}
\end{figure}

All measurements were performed on silicon MOSFETs. Such devices have been widely used because of the ability to continuously vary the electron density and the Fermi energy by use of a metal gate. The geometry of the device was chosen to be circular (Corbino geometry) to avoid leakage current paths around the source and drain contacts. The devices were fabricated using a high resistivity (10$^4$\,$\Omega$.cm) (100) p-silicon wafer to minimize, as much as possible, the scattering with boron acceptor impurities, especially close to the silicon-oxide interface. A 35\,nm thick gate thermal oxide was grown at 950\,$^{\circ}$C in a dry, chlorine-free oxygen atmosphere. The sidewalls of the oxide were protected by thick insulating LOCOS (Local Oxidation of Silicon). The effective gate length of the Corbino MOSFETs was 1\,$\mu$m and the diameter of the interior contact was 110\,$\mu$m. Contacts were realized by implanting phosphorous at high dose and sputtering aluminium. The contact resistivity was measured to be 3.5 and 2.3\,$\Omega$.cm$^{-1}$ at nitrogen and helium temperatures, respectively, and the sheet resistance was 6.3 and 5.9\,$\Omega$.$\Box^{-1}$ for the same temperatures. Sodium ions were introduced onto the oxide surface by immersing the device in a $10^{-7}$\,N solution ($\sim 6.4 \times 10^{11}\,$cm$^{-2}$) of high-purity sodium chloride in deionized water. The surface of the chip was then dried with nitrogen gas and an aluminium gate subsequently evaporated. The application of a positive gate voltage (+4\,V at $65^\circ$C for 10\,mins) causes the sodium ions to drift towards the Si-SiO$_2$ interface while the application of $-4$\,V DC in the same conditions removes the ions from the interface. The ions are frozen at their position once the device temperature becomes lower than approximately 150\,K (Fig. 1). Standard low-noise lock-in techniques with an amplifier gain of 10$^8$\,V/A were used to measure the source to drain conductivity. An AC excitation of amplitude $V_{\textup{\scriptsize{AC}}} = 15\,\mu$V and a frequency of 11\,Hz were chosen. The DC offset of the amplifier was suppressed using an appropriate blocking capacitor. The gate voltage was controlled by a high resolution digital to analog converter and the temperature measured by a calibrated germanium thermometer. The magnetic field was produced by an Oxford 12\,T superconducting magnet and applied perpendicular to the Si-SiO$_2$ interface.

Several devices were processed identically and gave results that lead to identical conclusions although we noticed some variation in the relative position and width of the impurity bands, as well as in the conductivity values. From the difference in the threshold voltage at 77\,K, we obtained an effective ion density of  $\sim 3.7 \times 10^{11}\,$cm$^{-2}$ at the interface indicating that only 60$\,\%$ of the ions drifted to the interface. We also fabricated a number of control devices that were not exposed to sodium contamination and were used for comparison. The following results are presented for a specific device that was chosen for its high reproducibility in time as well as for its high signal to noise ratio.

\section{Results and discussion}

Fig. 2a represents the source-drain conductivity obtained at different values of the magnetic field. The dependence of conductivity on temperature for the same device in the hopping regime showed the presence of two groups of peaks clustered around $V_{\textup{\scriptsize{g}}} = -2$ and $-0.5$\,V. These were attributed to a split impurity band due to the presence of sodium impurities at the Si-SiO$_2$ interface \cite{Ferrus1}. Arguments in favour of Hubbard bands were provided by studying the variation of conductivity at higher temperature, where activation mechanisms determine the behaviour \cite{Ferrus2}.

\subsection{Negative magnetoconductivity}

The magnetoconductivity was found to be negative for the whole range of gate voltages studied. This is expected and was already observed in localized systems \cite{Ghosh, Timp, Reisinger} (Fig. 2b). The variation with magnetic field is well described by ln\,$\sigma \sim -\alpha B^2$ even at fields up to 5\,T where $\alpha$ is gate voltage dependent. This behaviour is often attributed to an orbital compression of the donor wavefunctions due to the magnetic field. This results in a reduction of the wavefunction overlap between neighbouring hopping sites and an increase of the localization of electrons \cite{Shklovskii}. The wavefunction then acquires a phase factor $\phi \sim \left(\xi r^3 \right)^{1/2}/l_B^2$ where $l_B$ is the magnetic length given by $l_B = \left(\hbar/eB\right)^{1/2}$, $r$ is the hopping length and $\xi$ is the localization length. This assumes a gaussian distribution for the flux $\phi$. For transport processes involving only one hop within an area $\left( \xi r^3 \right)^{1/2}$ as in the case of nearest-neighbour hopping or resonant tunneling, ln\,$\sigma$ is found to be quadratic in magnetic field with $\alpha$ given by \cite{Nguen} :

\begin{eqnarray}\label{eqn:equation1}
\alpha\,=\,\frac{1}{12} \left(\frac{B_c}{\pi}\right)^{1/2} \frac{e^2 \xi} {\hbar^2 N_T^{3/2}}
\end{eqnarray}

$N_T$ is the density of active traps and $B_c$ is a constant describing the number of bonds in a circular percolation problem, being $\pi$ in tunnelling problems and respectively 4.5 and 2.7 in the case of nearest-neighbour hopping in a zero or finite width impurity band. If, however, the transport is described by several hops, a minimization procedure, as first used by Mott, \cite{Mott} needs to be carried out in order to find the optimum hopping length and hopping energy. This makes $\alpha$ temperature dependent \cite{Shklovskii2}.

\begin{eqnarray}\label{eqn:equation2}
\alpha\,=\,\frac{1}{24} \frac{e^2 \xi^4} {\hbar^2} \frac{1} {\left(p+3\right)^3}\left(\frac{T_0}{T}\right)^{3\left(\frac{p+1} {p+3}\right)}
\end{eqnarray}

where $p$ is 0 in case of the Mott variable range hopping regime \cite{Mott2} or 1 in case electron correlations are important (Efros and Shklovskii regime \cite{Shklovskii3})

\begin{figure}
\begin{center}
\includegraphics*[width=25pc]{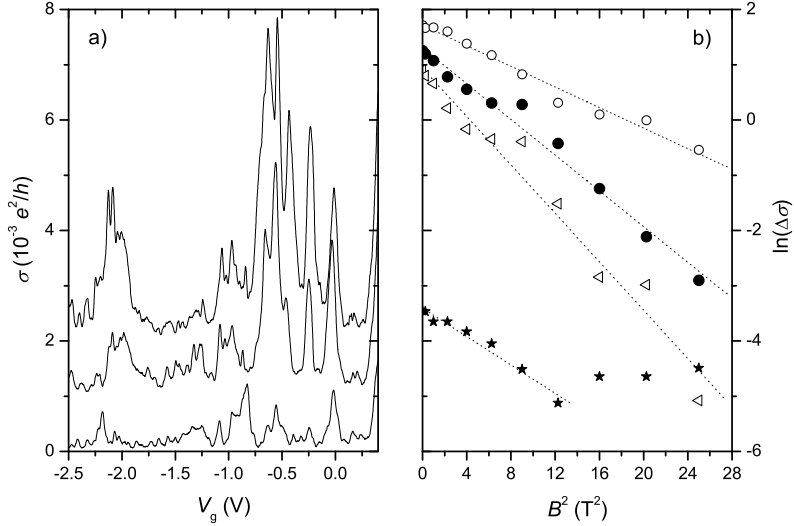}
\end{center}
\caption{a) Conductivity versus gate voltage at $B$\,=\,0\,T, 2.5\,T and 5\,T (from top to bottom) at 309\,mK. For clarity curves are separated by a constant offset of $10^{-3} e^2/h$. b) Magnetoconductivity for $V_{\textup{\scriptsize{g}}} = -0.24$ ($\circ$), $-0.56$ ($\bullet$), $-0.97$\,V ($\star$) and $-2.08$\,V ($\triangleleft$).}
\end{figure}

\begin{figure}
\begin{center}
\includegraphics*[width=33pc]{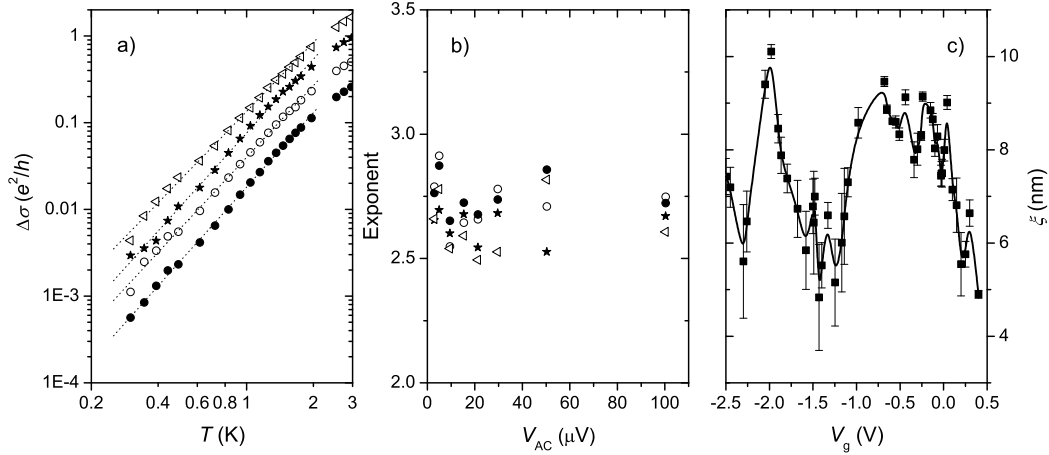}
\end{center}
\caption{a) Variation of the conductivity in temperature for $V_{\textup{\scriptsize{g}}} = -0.24$ ($\circ$), $-0.56$ ($\bullet$),$ -0.97$\,V ($\star$) and $-2.08$\,V ($\triangleleft$). b) Variation of the temperature exponent with $V_{\textup{\scriptsize{AC}}}$ at the same gate voltages. c) Variation of the localization length in terms of $V_{\textup{\scriptsize{g}}}$ at 309\,mK deduced from Eq. 2.}
\end{figure}

\subsection{Low temperature conductivity and localization length}

The proper analysis of $\sigma(B)$ is thus constrained to the correct description of the transport mechanism at low temperature. Below 1\,K, the conductivity decreases rapidly with temperature as $\Delta\sigma =\sigma\left(T\right)-\sigma\left(0\right) \sim\,T^{2.7}$ (Fig. 3a). This variation is not affected by the source-drain voltage and $eV_{\textup{\scriptsize{AC}}} < k_{\textup{\scriptsize{B}}}T$ in all measurements (Fig. 3b). Consequently, we do not expect that this behaviour results from electron heating in the contacts, but from a contribution of the acoustic phonons at low temperature. In our previous study, the same data was analyzed in temperature and we showed that the transport was characterized by hopping conduction with an exponent $p \sim 0.4$ above 1\,K \cite{Ferrus1}. In the framework of 2D resistance networks \cite{Miller} the conductivity is $\sigma \sim \nu/T$, whereas the scattering time is $\tau \sim \nu r^2/T$ where $\nu$ is the hopping rate and $r$ the hopping length. Our results showed the best fits were obtained with an exponential prefactor $\sim T^{-0.8}$. This implies $\tau$ varies as $T^{-1.6}$. This value is close to the $T^{-3/2}$ law expected in non-polar semiconductors for acoustic phonon scattering \cite{phonon} and as  observed in silicon \cite{phonon2}. This excellent agreement with theory implies that acoustic phonon-assisted hopping is the main mechanism for transport above 1\,K and that this transport is in fact still active and predominant at 300\,mK. Due to the short hopping distance, typically $r \sim\,3\xi$, an electron is scattered by only a few impurities between the initial and the final sites. Consequently, Eq. 2 should be used with $p = 0.4$ to estimate the value of the localization length $\xi$ at low temperature, giving the values of $T_0$ from our previous analysis \cite{Ferrus1}. Results are shown on Fig. 3c. The variation of $\xi\left(V_{\textup{\scriptsize{g}}}\right)$ is consistent with our previous observations, i.e. an increase in the localization at the band positions, but the values are now nearly half of those obtained from $T_0\left(\xi\right)$ in the absence of magnetic field \cite{Ferrus1}. Such a halving of $\xi$ has already been observed experimentally in insulating devices with strong spin-orbit coupling \cite{Pichard}. It is plausible that such an effect takes place in our device because of the presence of strong potential fluctuations at the Si-SiO$_2$ interface. We also observed that the conductivity of the lower band is suppressed at a field of about 3.5\,T whereas the upper band is still conducting even at the higher field. This suggests that the lower band is energically deeper and is already strongly localized even in the absence of a magnetic field. Effectively, the upper band is supposed to be closer to the extended states in energy than the lower band and it is likely to be more strongly influenced by the enhancement of the localization due to the shrinkage of the electron wave function caused by the magnetic field.
A point of interest is the absence of significant differences in the values of the localization length in the upper and lower bands. One may expect indeed the lower band to be more deeply localized and have a lower localization length. In fact, one needs to consider that, in our device and because of disorder, the lower Hubbard band may be formed by a majority of singly occupied states and a minority of empty or doubly-occupied states. In order to get transport through the lower Hubbard band, most of the electrons will hop to a singly-occupied state and form a D$_-$ state. On the other hand, in the upper Hubbard band, most of the sites will be doubly-occupied and the electrons will hop directly to the conduction band edge because of the presence of a large conduction band tail and the presence of states in the gap between the upper Hubbard band and the conduction band. As a consequence, we expect to find the same localization length for the upper and lower band in transport measurement in the hopping regime.

\subsection{Magnetoconductivity fluctuations}

We observed also a non-monotonic decrease in the conductivity with magnetic field. This takes the form of fluctuations with amplitude dependent on gate voltage. These fluctuations are present in many devices except for those without sodium ions. The fluctuations appear as a series of positive and negative magnetoconductivities. Their position in magnetic field also changes with time, but in general is periodic in $B$. Such fluctuations have already been observed by Nguyen \cite{Nguyen} and were attributed to random interference between different hopping paths. To procede with the analysis, the magnetic field value $B_{\textup{\scriptsize{m}}}$ of the fluctuation extrema were extracted from the curve $\Delta\sigma (B) = \sigma (B)-\sigma_0$ exp$\left(-\alpha B^2\right)$ and plotted as a function of the extrema index (Fig. 4). We found a linear dependence with a slope $B_{\phi}$ for all gate voltages indicating the fluctuations are periodic in magnetic field. To minimize the uncertainty due to the limited number of points, the experiment was repeated four times with the same configuration. We found that $\xi_{\phi} = {\left(\hbar/eB_{\phi}\right)}^{1/2} > \xi$ but $\xi_{\phi} \sim \left(r\xi\right)^{1/2}$. We can deduce the modulation corresponds to an integer number of elementary flux quanta through a surface of area $\pi r \xi$ and that $B_{\textup{\scriptsize{m}}}$ is given by :

\begin{eqnarray}\label{eqn:equation3}
B_{\textup{\scriptsize{m}}}\,=\,\frac{nh}{2\pi e r \xi}
\end{eqnarray}

\begin{figure}
\begin{center}
\includegraphics*[width=25pc]{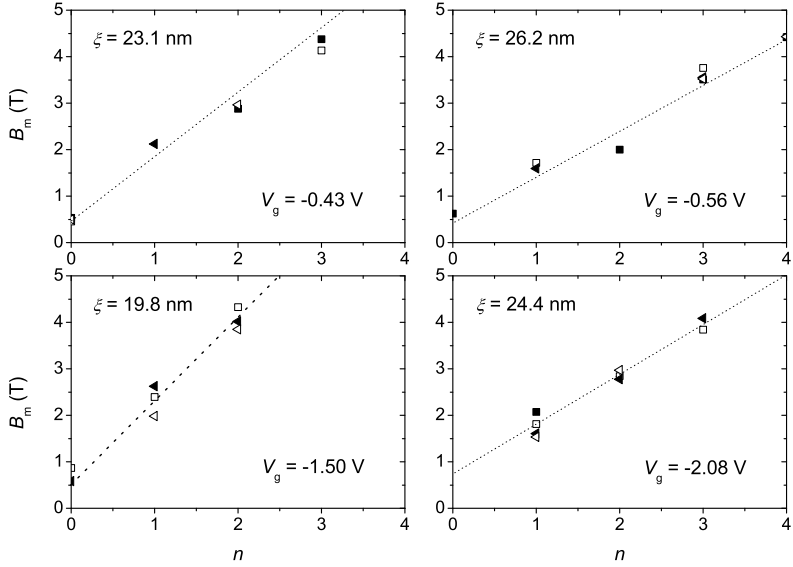}
\end{center}
\caption{Measure of the critical field of the fluctuations and the corresponding values $\xi$ for $V_{\textup{\scriptsize{g}}} = -0.43$\,V, $-0.56$\,V,$ -1.50$\,V and $-2.08$\,V at 309\,mK.}
\end{figure}

This situation is similar to that giving rise to Aharonov-Bohm oscillations in quantum dot systems \cite{Aharonov} with $\left(r\xi\right)^{1/2}$ playing the role of the dot radius. In the hopping regime, the interference area has a cigar-shape with radius $\xi$ in one direction and the hopping length in the hop direction. Quantum interferences between forward and backscattering paths induced by the localizing impurity potential due to the magnetic field may be responsible for such fluctuations.

\section{Peak positions in magnetic field}

Finally, we discuss the variation in the position of the peaks in magnetic field (Fig. 5). We observed that the group of peaks clustered in the upper band moves towards more negative gate voltages with approximately a slope of $\sim -4\,$mV/T whereas the one in the lower band move with a slope of $\sim +4\,$mV/T. Such movement in energy levels are expected in magnetic field, and the difference in the sign can be explained by including the effect of the orbital momentum. In fact, this behaviour cannot be interpreted as being due to spin effects as no spin splitting was observed. Also, the gap energy between the upper and lower bands estimated previously \cite{Ferrus2} is much larger than the spin splitting at 5\,T. For simplicity, we suppose the electrons are localized at the impurity sites with an approximate parabolic potential of width $\omega_0$. The variation $\Delta E$ of an energy level is then described by \cite{Fock, Darwin}:

\begin{eqnarray}\label{eqn:equation4}
\Delta E\,=\,\left(2n + |l| + 1\right) \hbar \left[\sqrt{\frac{\omega_c^2}{4}+\omega_0^2}-\omega_0 \right] - \frac{1}{2} l \hbar\omega_c
\end{eqnarray}

where $n$ is the index for the Landau levels, $l$ is the value of the angular momentum and $\omega_c = eB/m^*$ with $m^* = 0.19\,m_{\scriptsize{\textup{e}-}}$ the transverse effective mass of the electrons in $<100>$ silicon.

\begin{figure}
\begin{center}
\includegraphics*[width=25pc]{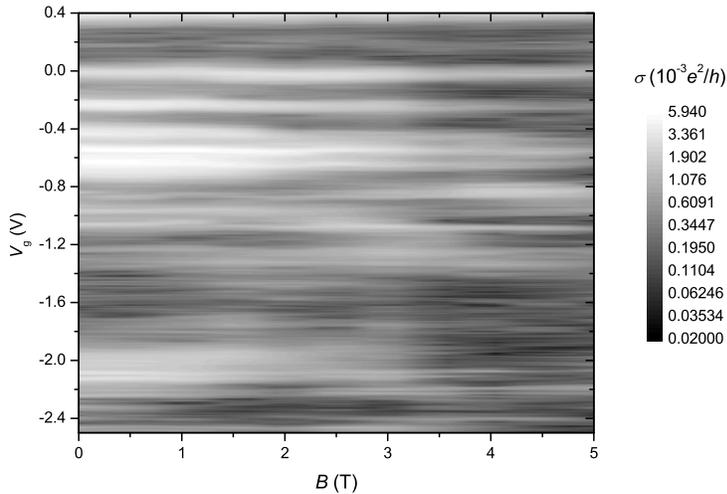}
\end{center}
\caption{Conductivity versus gate voltage and magnetic field at 309\,mK.}
\end{figure}

Consequently, $n = 0$ and $l = 0$ are associated with the lower group of peaks (non-degenerate) and $n = 1$ and $l = 1$ to the upper group (degeneracy of 2). The value $l = 1$ also implies the presence of two electrons in the configuration 1s2p. This behaviour supports the idea of the formation of Hubbard bands \cite{Hubbard2} and
a stable Na$^-$ ion at the Si-SiO$_2$ interface as well as the presence of a strong spin-orbit coupling. This shows as well the importance of Coulomb interactions in this system. When increasing the magnetic field, the wave-function of the lowest state shrinks and the interaction between the two degenerate electrons grows until the Zeeman energy is sufficient to spin-flip the second electron and promote it to the $l = 1$ state. This transition is predominantly caused by Coulomb interactions since no spin splitting was observed and was still quite small even at 5\,T. Finally, the 1s2p configuration holds as long as $\Delta E < 0$ for the upper band, which is verified up to 5\,T. This implies that $\hbar \omega_0 > 2.7\,$meV and provides a lower bound for the mean impurity potential depth.

\section{Conclusion}

We have observed a negative magneto-conductivity up to 5\,T. This result confirms previous observations on the same device that the system is strongly localized with a localization length of the order of the mean distance between impurities. The fluctuations observed in the magnetoconductivity have been attributed to quantum interference due to the presence of impurity centres close to the interface. Finally, the shift in the position of the bands are compatible with the presence of upper and lower Hubbard bands and a possible transition from a singlet to a triplet state in the upper band induced by electron-electron interaction.

\section*{Acknowledgement}

We would like to thank Drs T. Bouchet and F. Torregrossa from Ion Beam Services-France for the process in the device as well as funding from the U.S. ARDA through U.S. ARO grant number DAAD19-01-1-0552.


\begin{thebibliography}{31}

\bibitem{Fang} F.F. Fang, A.B. Fowler, Phys. Rev. 169 (1967) 619.

\bibitem{Hartstein} A. Hartstein, A.B. Fowler, Phys. Rev. Lett. 34 (1975) 1435.

\bibitem{Pepper} M. Pepper, J. Phys. C : Solid State Phys. 10 (1977) L173.

\bibitem{Suton} A.P. Sutton, M.W. Finnis, D.G. Pettifor, Y. Ohta , J. Phys. C : Solid State Phys. 21 (1988) 35.

\bibitem{Hubbard} J. Hubbard, Proc. Roy. Soc. A 276 (1963) 238.

\bibitem{Norton} P. Norton, Phys. Rev. Lett. 37 (1976) 164.

\bibitem{Taniguchi} M. Taniguchi, S. Narita, Solid State Comm. 20 (1976) 131.

\bibitem{UHB} A.B. Fowler, J.J. Wainer, R.A. Webb, IBM J. Res. Develop. 32 (1988) 372.

\bibitem{Kane} B.E. Kane, Nature 393 (1998) 133.

\bibitem{Kishida} H. Kishida, H. Matsuzaki, H. Okamoto, T. Manabe, M. Yamashita, Y. Taguchi, Y. Tokura, Nature 405 (2000) 929.

\bibitem{Wu} D. Wu, Y.C. Ma, Z.L. Wang, Q. Luo, C.Z. Gu, N.L. Wang, C.Y. Li, X.Y. Lu, Z.S. Jin, Phys. Rev. B 73 (2006) 012501.

\bibitem{Ferrus1} T. Ferrus, R. George, C.H.W. Barnes, N. Lumpkin, D.J. Paul, M. Pepper, Phys. Rev. B 73 (2006) 041304(R).

\bibitem{Ferrus2} T. Ferrus, R. George, C.H.W. Barnes, N. Lumpkin, D.J. Paul, M. Pepper, J. Phys. : Condens. Matter 19 (2007) 226216.

\bibitem{Ghosh} A. Ghosh, M. Pepper, H.E. Beere, D.A. Ritchie, Phys. Rev. B 70 (2004) 233309.

\bibitem{Timp} G. Timp, A.B. Fowler, Phys. Rev. B 33 (1986) 4392.

\bibitem{Reisinger} H. Reisinger, A.B. Fowler, A. Hartstein, Surf. Sci. 142 (1984) 274.

\bibitem{Shklovskii} B.I. Shklovskii, A.L. Efros, in: Electronic properties of doped semiconductors, Springer Series in Solid-State Sciences, vol 45, Springer-Verlag, Berlin (1984).

\bibitem{Nguen} V.L. Nguen, Sov. Phys. Semicond 18 (2) (1984) 207.

\bibitem{Mott} N.H. Mott, J. Non-Cryst. Solids 1 (1968) 1.

\bibitem{Shklovskii2} B.I.  Shklovskii, Fiz. Tekh Poluprovodn. (S.-Petersburg) 17 (1983) 2055 [Sov. Phys. Semicond. 17 (1983) 1311].

\bibitem{Mott2} N.F. Mott, E.A. Davis, Electronic Processes in Non-Crystalline Materials, 2nd ed., Oxford University Press, London (1979).

\bibitem{Shklovskii3} A.L. Efros, B.I. Shklovskii, J. Phys. C: Solid State Phys. 8 (1975) L49.

\bibitem{Miller}  A. Miller, E. Abrahams, Phys. Rev. 120 (1960) 745.

\bibitem{phonon} J. Bardeen, W. Schockley, Phys. Rev. 80 (1950) 72.

\bibitem{phonon2} M.B. Prince, Phys. Rev. 93 (1954) 1204.

\bibitem{Pichard} J.L. Pichard, M. Sanquer, K. Slevin, P. Debray, Phys. Rev. Lett. 65 (1990) 1812.

\bibitem{Nguyen} V.L. Nguyen, B.Z. Spivak, B.I. Shklovskii, Zh. \'Eksp. Teor. Fiz. 89 (1985) 1770 [Sov. Phys. JETP 62 (1985) 1021].

\bibitem{Aharonov} R.A. Webb, S. Washburn, C.P. Umbach, R.B. Laibowitz, Phys. Rev. Lett. 54 (1988) 2968.

\bibitem{Fock} V. Fock, Z. Phys. 47 (1928) 446.

\bibitem{Darwin} C.G. Darwin, Proc. Cambridge Philos. Soc. 27 (1930) 86.

\bibitem{Hubbard2} P. Norton, Phys. Rev. Lett. 37 (1976) 164.


\end{thebibliography}
\end{document}